\documentclass[iop,twocolappendix,numberedappendix]{emulateapj}

\usepackage{amsfonts,amsmath,units,wasysym,epsfig,graphicx,verbatim,color,subfigure,graphicx}
\usepackage{amsmath}
\usepackage{latexsym}
\usepackage{amssymb}
\usepackage{amsfonts}
\usepackage{mathtools}
\usepackage{bm}
\usepackage{color}
\usepackage{float}
\usepackage{tikz}
\usepackage{graphicx}
\usepackage{journals}
\usepackage[breaklinks,colorlinks,citecolor=blue]{hyperref}
\usepackage{natbib}

\def\be{\begin{equation}}
\def\ee{\end{equation}}
\def\bea{\begin{eqnarray}}
\def\eea{\end{eqnarray}}

\begin{document}

\newcommand{\gpch}{h^{-1}{\rm Gpc}}
\newcommand{\mpch}{h^{-1}{\rm Mpc}}
\newcommand{\kmspMpc}{~{\mathrm {kms}}^{-1}{\mathrm {Mpc}}^{-1}}
\newcommand{\msunh}{h^{-1}M_\odot}
\newcommand{\IUCAA}{Inter-University Centre for Astronomy and
  Astrophysics, Post Bag 4, Ganeshkhind, Pune 411 007, India}
\newcommand{\IPMU}{Kavli Institute for the Physics and Mathematics of the Universe (WPI)
  , 5-1-5, Kashiwanoha, 277-8583, Japan}
\newcommand{\WSU}{Department of Physics \& Astronomy, Washington State University, 1245 Webster, Pullman, WA 99164-2814, U.S.A.}

\newcommand{\todo}[1]{{\textcolor{red}{\bf{Todo: #1}} }}

\newcommand{\sukanta}[1]{{\textcolor{magenta}{\bf{#1}} }}

\date{\today}

\title{Incompleteness matters not: Inference of $H_0$ from BBH-galaxy cross-correlations}

\author{Sayantani Bera$^1$}
\email{sayantani@iucaa.in}

\author{Divya Rana$^1$}

\author{Surhud More$^{1, 2}$}


\author{Sukanta Bose$^{1, 3}$}
\affiliation{$^1$\IUCAA}
\affiliation{$^2$\IPMU}
\affiliation{$^3$\WSU}

\begin{abstract}

We show how the angular clustering between gravitational-wave standard sirens and galaxies with known redshifts allows an inference of the Hubble constant, regardless of whether the host galaxies of any of these sirens are present in the galaxy catalog. We demonstrate this for the first time with realistic simulations of gravitational-wave signals from binary black holes in a three-detector network with Advanced LIGO and Advanced Virgo sensitivities. We show that with such a network, the cross-correlation technique can be used to infer the Hubble parameter with a precision of less than 10\% (2\%) at 90\% confidence with 50 (500) sources, even with a 100\% {\it incomplete} catalog, which does not contain the hosts of any of the gravitational-wave events. We compare our method with the current state-of-the-art techniques used for the inference of the Hubble parameter from real data. We argue that, if the clustering information is not used explicitly, the inference of $H_0$ from real data is expected to be prior-dominated.

\end{abstract}

\section{Introduction}
\label{intro}

The constant of proportionality in the Hubble-Lemaitre expansion law of the Universe, called the Hubble constant, is an important parameter in the concordance cosmological model \citep{1929PNAS...15..168H,1931MNRAS..91..483L}. The Hubble constant sets the current rate of cosmological expansion by relating the observed redshift of galaxies to their distances. It enters all distance and time measurements in the Universe \citep[see, e.g.,][]{Hogg99}. Therefore, it is critically important that measurements of the Hubble constant are not only precise but also accurate.

The determination of the Hubble constant has traditionally relied on the distance ladder in the local Universe. A variety of distance probes utilizing geometric methods such as parallaxes and mega-maser observations in addition to standard candles such as Cepheid variables and supernovae of Type Ia, that are applicable at different distances, are calibrated against each other at every rung of the distance ladder \citep{2007LRR....10....4J}. The best measurements of the Hubble constant using the distance ladder technique comes from the Supernovae, $H_0$, for the Equation of State of Dark energy (SHoES) project, which reported a measurement of the Hubble constant with an accuracy of order 2 percent \citep{2016ApJ...826...56R,2018ApJ...855..136R,2018ApJ...861..126R,2019ApJ...876...85R,2019ApJ...886L..27R}.

The cosmic microwave background (CMB) experiment Planck provides an independent measurement of the Hubble constant by observing the sound horizon in the very early Universe \citep{Ade:2015xua,Planck2018}. A reverse distance ladder utilizing the CMB measurements and the measurements of the Baryon acoustic oscillations provide yet another way of inferring the Hubble constant \citep{2018ApJ...853..119A, 2019JCAP...10..029S,2020JCAP...05..032P}. The values of the measurements obtained from the distant Universe are at odds with the values inferred from the local Universe distance ladder at a significance of about $\sim4\,\sigma$ \citep{2019ApJ...876...85R, Verde:2019ivm}. The difference in the two measurements is an uncomfortable conundrum for the concordance cosmological model. If systematic effects could be ruled out, this difference between the two measurements would indicate a breakdown of the concordance cosmological model, or a possible new physics in the matter sector \citep{PhysRevD.94.103523,PhysRevD.97.043528,2018JCAP...05..052N,2018JCAP...09..025M,2019PhRvD..99j3526K,2019PhRvL.122v1301P,2020arXiv200406114B,Kreisch:2019yzn,Lin:2019qug,Knox:2019rjx}.

The simultaneous detection of gravitational wave event GW170817, resulting from a binary neutron star merger, and the electromagnetic radiation in the form of $\gamma$ rays from this event has ushered in a new era for multi-messenger astronomy \citep{TheLIGOScientific:2017qsa}. The gravitational waves act as standard sirens, the amplitude of the detected waves in combination with the characteristics of the waveform (such as the rate of change of frequency) provides a measurement of the luminosity distance to the event \citep[see e.g.,][]{ Schutz:1986gp,Holz:2005df}. The successful search for the galaxy that acted as the host of the event enables a measurement of the redshift, thus providing an avenue to infer the Hubble constant in the local Universe independent of the distance ladder \citep{Soares-Santos:2019irc,Abbott:2017xzu}. This has raised the prospect of gravitational wave observations acting as the arbiter for determining the Hubble constant in the local Universe. Unfortunately, in the three runs of the LIGO-Virgo gravitational wave network that have concluded so far, this is the sole gravitational wave event that has been simultaneously observed in the electromagnetic spectrum conclusively with several observatories~\citep{LIGOScientific:2018mvr,LIGOScientific:2020Sci}. The large observational follow-up campaign enabled ever improving measurement of $H_0$, which most recently is estimated to be
$68^{+14}_{-7}\kmspMpc$ (at 68.3\% highest-density posterior interval with a flat-in-log prior) after combining various LIGO-Virgo observations from their first and second observation runs, including the binary neutron star merger GW170817 with a precisely measured redshift owing to the identification of its electromagnetic counterpart and host galaxy~\citep{Abbott:2017xzu,Abbott:2019yzh,Soares-Santos:2019irc}.

The vast majority of gravitational wave (GW) events detected thus far, however, correspond to the mergers of massive binary black holes, which outnumber the binary neutron star events by about an order of magnitude. There is little evidence so far that such events have electromagnetic counterparts~\citep{LIGOScientific:2018mvr}, pursuing which  is admittedly non-trivial~\citep{Aasi:2013wya,Rana:2016crg,Rana:2019uti}. There have been ongoing efforts towards extracting cosmological information from such observations for which the redshift information is unavailable \citep{MacLeod:2007jd,DelPozzo:2012zz,Chen:2017rfc,Nair:2018ign,Fishbach:2018gjp,Ding:2018zrk,Vitale:2018wlg,Borhanian:2020vyr}. Of particular relevance is \citet{MacLeod:2007jd}, which introduced the possibility of using galaxy clustering information and the galaxy redshift distribution, albeit in regards to possible sources for GW detectors in space, and \citet{Oguri:2016dgk} in the context of the current and next generation gravitational wave detectors.

The idea of a joint examination of such gravitational wave events with galaxy catalogs with known redshifts to determine the Hubble constant was first presented in \citet{Schutz:1986gp}. Large redshift surveys of galaxies had shown that the spatial distribution of bright galaxies is clustered \citep{1982ApJ...253..423D,1983ApJS...52...89H,Geller:1989da, 1999ApJS..121..287H,1999PASP..111..438F,1978AJ.....83.1549K,1981ApJ...248L..57K,2002AJ....123..485S,2006ApJS..162...38A,2005MNRAS.358..441B,2002AJ....124.1810S}. This clustering allows a statistical inference of the redshift of the gravitational wave event, and of the Hubble constant, even in the absence of any knowledge of the true host of any gravitational wave event. In this paper, we present a practical application of this method to infer the redshift distribution of gravitational wave events arising from the coalescence of binary black holes (BBHs), without the need for explicitly associating galaxies with known redshifts in a catalog as potential hosts of such events. It is worth mentioning that that the clustering measurement can also facilitate the inference of the possible astrophysical origin of the BBH mergers \citep{Adhikari:2020wpn}.

After the successful O1 and O2 observation runs of LIGO in the past years, the detection of close to a dozen BBH signals has enabled a successful implementation of a variation of Schutz's method on real events \citep{Abbott:2019yzh,Soares-Santos:2019irc,Fishbach:2018gjp, Abbott:2020khf,Palmese:2020aof}. Although the constraints obtained in this way are much weaker compared to that obtained from the BNS GW170817 event, it nonetheless gives an independent method of recovering the $H_0$ distribution where no redshift information about the GW event is available. The LIGO analysis paper \citep{Abbott:2019yzh} closely follows the analysis method described in~\citep{Gray:2019ksv}. In \citet{Gray:2019ksv}, the authors adopt a Bayesian inference method for parameter estimation and demonstrate the method using simulated galaxy catalogs with varying degree of completeness.\footnote{Here completeness refers to the fraction of gravitational wave events whose hosts are part of the galaxy catalogs.} Their analysis demonstrates that a few hundred gravitational wave events can be used to constrain the Hubble constant with 4.4\% percent accuracy even when a mere 50\% of the events in the sample have a host in their galaxy catalog.

However, the Bayesian method adopted in their analysis assumes no underlying clustering of galaxies at all, and was correspondingly tested on mock catalogs where galaxies and the potential hosts of the gravitational wave events were distributed in a random manner in the simulated comoving volume. The Bayesian method thus accounts for two possible cases, based on whether the true host is (a) in the catalog or (b) or outside the catalog. For a complete galaxy catalog, the contribution comes solely from the first term for which the sky positions and redshift distributions are well known (typically given by either a Dirac $\delta$-function or a Gaussian distribution), thus giving a much tighter constraint. As the incompleteness increases, the contribution from the second term (host outside the catalog) increases significantly. Since there is no information about the sky localizations or redshift of the galaxies that have not been catalogued, this results in huge uncertainties, showing up as weaker constraints in the final analysis of the $H_0$ distribution. We show in Sec.~\ref{sec2a}, that in the limit of total incompleteness (i.e., when none of the galaxies in the catalog is
a host of any of the gravitational-wave events), the posterior distribution for the Hubble constant, given the data, will have to be entirely dominated by the priors. Thus, although this method seems functionally similar to the method proposed by \citet{Schutz:1986gp}, it does differ in spirit since it does not account for the clustering of galaxies.

In this work we demonstrate, with simulated signal injections and noise of a network of two or three of the Advanced LIGO \citep{TheLIGOScientific:2014jea} and Virgo detectors \citep{TheVirgo:2014hva}, that once clustering of galaxies is taken into account one can measure $H_0$ even when none of the gravitational-wave events has its host in the galaxy catalog. We argue that our method should be statistically more powerful than present methods as it does not rely on the single specific host of an event being present in the catalog, but relies on the clustering of a larger number of galaxies, which in turn are expected to be clustered with the gravitational wave events.

\section{BBH mergers as standard sirens}
\label{sec2}

Binary black hole (BBH) mergers are the most numerous events detected by LIGO and Virgo. Their gravitational wave signals provide information about their luminosity distance. In the absence of electromagnetic counterparts or any other means of determining  their redshifts -- or that of their host galaxies -- the inference of the Hubble constant from their observations necessitates an approach that is different from the case of binary neutron star mergers whose redshifts can be ascertained~\footnote{This worked for GW170817 but not for GW190425~\citep{Abbott:2020uma}, which has not yet been associated with an EM counterpart or a host galaxy.}. Under the assumption that binary black hole mergers would occur in galaxies, the current state of the art method for the inference of the Hubble constant relies on galaxy catalogs with sky position and redshift information (either spectroscopic or good photometric redshifts). When the catalog is deep enough, the chances of finding the host galaxy of a BBH merger in it improve.

\subsection{GW event-galaxy host association:  State of the art}
\label{sec2a}

The formal posterior distribution of the Hubble constant ($H_0$) given the BBH merger event detections (${D_{\rm GW}}$), their time series data (${x_{\rm GW}}$) and the input galaxy catalogs in the localization regions of these events can be written down as \citep{Gray:2019ksv}
\begin{eqnarray}
P(H_0|\{D_{\rm GW}\},\{x_{\rm GW}\}) &\propto& P(H_0)P(N_{\rm det}|H_0)\nonumber \\ 
& \times& \prod_{i}^{N_{\rm det}} p(x_{{\rm GW},i}|D_{{\rm GW},i}, H_0)\,,\nonumber\\
\end{eqnarray}
where $P(H_0)$ is the prior distribution, $P(N_{\rm det}|H_0)$ is the probability of detecting $N_{\rm det}$ events given a value of $H_0$, while the remaining term is the product over the likelihoods of the detector time-series for each event, with index $i$.

Information about the Hubble constant is obtained by associating each individual event with the known redshift distribution of the galaxies in the input catalog that are in the localization region. 
For a detection
that corresponds to a galaxy existing in the catalog (denoted by $G$), this probability $P(x_{\rm GW}| G, D_{\rm GW}, H_0)$ is a sum over likelihoods of the detector time-series constructed by assuming each galaxy to be its potential host. This likelihood accounts for the position of the galaxy compared to the event localization and its amplitude given the value of its redshift and the Hubble constant \citep[see Eq.~A.11 in][]{Gray:2019ksv}. As the number of events increases, statistically the influence of the non-host galaxies (in the foreground and background) on the posterior of $H_0$ will grow slower than that of the true hosts, roughly by a factor of $\sqrt{N_\mathrm{det}}$.

Currently, no observational clues about the nature of the hosts of the BBH mergers exist apart from educated guesses. Empirically it is unclear if such events occur preferentially in more massive galaxies~\footnote{The components of BBH events could even be primordial black holes, in which case we do not expect a galaxy to be the host~\citep{1971MNRAS.152...75H,1974MNRAS.168..399C,Carr:1975qj,Raidal:2017mfl}.}. Thus, the determination of whether galaxy catalogs themselves are deep enough for the current clutch of BBH events has considerable uncertainty. Consequently, the posterior calculation of the Hubble constant must account for the possibility that the host galaxy of any given BBH does not exist in the catalog ($\tilde{G}$). This is effected by weighting the likelihood of each BBH with the probability that its host galaxy is in the set $G$ and combining it with a similar term representing the alternative scenario (that the host is in set $\tilde{G}$). The latter term is expected to marginalize over all possible unseen galaxies -- with very uncertain redshift distributions and with no sky position information \citep[see Eq.~A.21 in][]{Gray:2019ksv}. This causes the posterior distribution of the Hubble constant to be dominated by the assumed priors instead of the data itself.

Combining these two types of terms, the individual likelihoods become
\begin{multline}
p(x_{{\rm GW}} | D_{{\rm GW}}, H_0)\\ 
    =\sum_{g=G, \tilde{G}} p(x_{{\rm GW}}|g, D_{{\rm GW}}, H_0) p(g|D_{\rm GW}, H_0)\,,
\end{multline}
where $p(g|D_{\rm GW}, H_0)$ denotes the probability that the galaxy exists in the catalog ($g = G$) or not ($g = \tilde{G}$).
In the most extreme case of $100\%$ incompleteness (i.e., where none of the host galaxies are present in the galaxy catalog), the posterior thus will be entirely dominated by the priors.

It is known, however, that the BBH events will be clustered with respect to the galaxy distribution in the Universe. Thus even though the galaxies in the input galaxy catalog are not the hosts of the BBH event, the galaxies at a similar redshift as the event will be clustered with respect to the BBH event. These galaxies will nevertheless contribute information that is not included in the formal posterior distribution as written out by \citet{Gray:2019ksv}.

In this paper, we simulate BBH events at locations determined by randomly sampling low-mass halos \footnote{The halo mass was chosen to have a bias equal to unity, thus making this sampling equivalent of randomly sampling matter particles.} from a cosmological N-body simulation that follows the evolution of large scale structure. The input galaxy catalog that we use to cross-correlate corresponds to a higher threshold on the halo masses. Therefore, we study the extreme case where our input galaxy catalog has $100\%$ incompleteness. In our case, by construction, we have
\begin{equation}
    p(G|D_{\rm GW}, H_0) = 1-p(\tilde{G}|D_{\rm GW}, H_0) = 0\,.
\end{equation}
Thus, the posterior distribution of the Hubble constant would be entirely dominated by the priors, with the data providing very little information. However, we show that the clustering information of the gravitational wave sources with galaxies in the input galaxy catalog (regardless of the presence of absence of the host galaxy in the catalog) can provide information about the Hubble constant. A clear demonstration of this effect is the main goal of this paper. We will defer the inclusion of the clustering information in the Bayesian formalism to a future work~\citep{Bera:inprep}.

\subsection{Redshift distribution of GW events using cross-correlation with galaxies}
\label{sec2b}

The true redshift distribution of a sample of astrophysical sources of unknown redshifts -- but sources that are expected to trace the large scale structure in the Universe -- can be recovered by statistically 
cross-correlating their positions with objects whose redshifts are known~\citep{Newman:2008mb}. The redshift distribution of a photometric sample of galaxies used for weak gravitational lensing is often calibrated from the angular two-point correlation with a spectroscopic sample with known redshift \citep{McQuinnWhite:2013, Menard:2013, Schmidt:2013, Mubdi:2015, Oguri:2016dgk, Johnson:2017, Zhang:2018ekk}.

To understand how large scale structure information can be used to recover the redshift distribution of a population of GW sources, consider the three-dimensional cross-correlation function $\xi_{\rm gw, g}(r)$ of a galaxy population with a set of GW events both of which trace the same large scale structure,
\begin{equation}
\xi_{\rm gw,g}(r) = \langle \delta_{\rm gw}(r)\delta_{\rm g}(r) \rangle\,.
\label{eqn:xi}
\end{equation}
Here, $r$ is the comoving distance between a GW source and galaxy pair. The functions $\delta_{\rm gw}$ and $\delta_{\rm g}$ are the number density contrasts of the GW source distribution and the galaxy distribution, respectively. Since both the GW source population and the galaxy population trace the same underlying matter distribution
, in the linear regime these density contrasts are related to the matter density contrast $\delta_{\rm m}$ as
\begin{eqnarray}
\delta_{\rm gw} &=& b_{\rm gw} \delta_{\rm m} \,, \\ 
\delta_{\rm g} &=& b_{\rm g} \delta_{\rm m} \,.
\end{eqnarray}
Here the proportionality constants $b_{\rm gw}$ and $b_{\rm g}$ are the linear biases of the population of GW sources and galaxies, with respect to the matter distribution, respectively. In general, the bias can be a function of different properties of the sources. A bias of unity would mean that the observed population is randomly sampled from the underlying matter distribution.\\

Expanding Eq.~\eqref{eqn:xi}, the 3D cross-correlation function $\xi_{{\rm gw,g}}(r)$ can be expressed in terms of the volume number densities of the galaxy and GW source populations $\bar{n}^{vol}_{{\rm g}}$ and $\bar{n}^{vol}_{{\rm gw}}$, respectively. If $n^{vol}_{{\rm gw,g}}(r)$ is the number density of galaxy-GW source pairs separated by a comoving distance $r$, then we can write
\begin{equation}
n^{vol}_{{\rm gw,g}}(r) = \bar{n}^{vol}_{{\rm gw}}\bar{n}^{vol}_{{\rm g}}[1+\xi_{{\rm gw,g}}(r)] 4\pi r^2 dr \,.
\end{equation}
If we consider galaxies at a redshift $z$ then the above correlation function becomes a function of $z$.
\begin{equation}
n^{vol}_{{\rm gw,g}}(r,z) = \bar{n}^{vol}_{{\rm gw}}\bar{n}^{vol}_{\rm g}(z)[1+\xi_{{\rm gw,g}}(r,z)] 4\pi r^2 dr\,,
\end{equation}
where the relevance of $\xi_{{\rm gw,g}}$ to observables we utilize in our method for measuring $H_0$ will become clear below.

Most of the time, it is convenient to work with two-dimensional number densities (i.e., surface number densities) and the corresponding 2D angular cross-correlations $w(\theta)$ when the full three-dimensional information is not readily available. The surface density of galaxy-gravitational wave source pairs, $n_{\rm gw,g}(\theta, z)$, where the galaxy within the pair is at a redshift $z$ and the distance between the pair lies between $\theta\pm d\theta/2$ is given by
\begin{equation}
    n_{\rm gw,g}(\theta, z) = \bar{n}_{\rm gw}\bar{n}_{\rm gal}(z)\left[ 1 + w_{\rm gw,g}(\theta)\right]\,2 \pi \theta d\theta \,.
\end{equation}
In the above equation, $\bar{n}_{\rm gw}$ and $\bar{n}_{\rm gal}(z)$ represent the surface number densities of the gravitational wave sources and the galaxies, respectively, and $w_{\rm gw,g}$ denotes the angular cross-correlation between the two populations. The angular cross-correlation function is a line-of-sight integral over the three-dimensional cross-correlation function, $\xi_{\rm gw, g}$ and satisfies the relation \citep{Newman:2008mb}
\begin{equation}
    w_{\rm gw, g}(\theta, z) = \int \Phi_{\rm gw}(z') \xi_{\rm gw, g}(r[\theta, z, z'], z) dz'\,,
\end{equation}
where $\Phi_{\rm gw}(z')$ denotes the redshift distribution of the gravitational wave sources, and $r$ is the aforementioned three-dimensional comoving distance between the galaxy and the gravitational-wave source, which lie at redshifts $z$ and $z'$, respectively, and are separated by angle $\theta$.

Given that the three dimensional correlation function, $\xi_{\rm gw, g}(r, z)$ falls rapidly with distance, $r$, only a fraction $\Phi_{\rm gw}(z')$ of sources that lie at redshifts very similar to that of the galaxies that we cross-correlate with, will contribute to the clustering signal, $w_{\rm gw, g}(\theta, z)$. Integrating the above equation over $\theta\in[0, \theta_{\rm max}]$, and assuming that the three dimensional cross-correlation function $\xi_{\rm gw, g}$ is given by a power law $(r/r_0)^{-\gamma}$, one obtains
\begin{equation}
    \Phi_{\rm gw}(z) = w(\leq \theta_{\rm max}, z) \left[ \frac{ f(\gamma) {d\chi}/{dz} }{r_0^{\gamma}d_{\rm A}^{1-\gamma}\theta_{\rm max}^{3-\gamma}}\right]\,, 
\end{equation}
where $\chi$ is the comoving distance and $d_{\rm A}(z)$ is the angular diameter distance at redshift $z$. The factor $f(\gamma)$ is given by,
\begin{equation}
    f(\gamma) = \frac{(3-\gamma)}{2\pi} \frac{\Gamma(\gamma/2)}{\Gamma(1/2)\Gamma([\gamma-1]/2)}\,,
\end{equation}
where $\Gamma(x)$ denotes the Gamma function. Note that even though there are mild dependencies on cosmology on the right-hand side, it is entirely free of the Hubble constant. 

We have tested that the factors inside the square brackets introduce fairly mild dependence on the redshift and, consequently, the mean redshift of $\Phi_{\rm gw}(z)$ can be determined from the dependence of $w(\leq \theta_{\rm max}, z)$ on the redshift~\footnote{We used the true redshift distribution of the GW sources in the mock Universe and multiplied it with the redshift-dependent factors in the square brackets. We found that the mean redshift of the product does not change significantly to affect our conclusions.}. In particular, $r_0$ depends on the clustering bias of the galaxy population as well as the gravitational wave population with respect to the matter distribution in the Universe. The dependence of galaxy bias on the redshift can be informed by measurements of the auto-correlation functions of galaxies in a given redshift slice with themselves. The bias of gravitational wave sources will have to assume a flexible functional form and marginalized over~\citep[see e.g.,][]{Mukherjee:2020}. In our simulations, we will not include a redshift dependence of clustering bias for galaxies or gravitational wave sources. Applications to real data should, however, include a proper modelling of these terms~\citep{Bera:inprep}.

\begin{figure*}
\centering
 \includegraphics[width = \textwidth, height=0.67\textwidth]{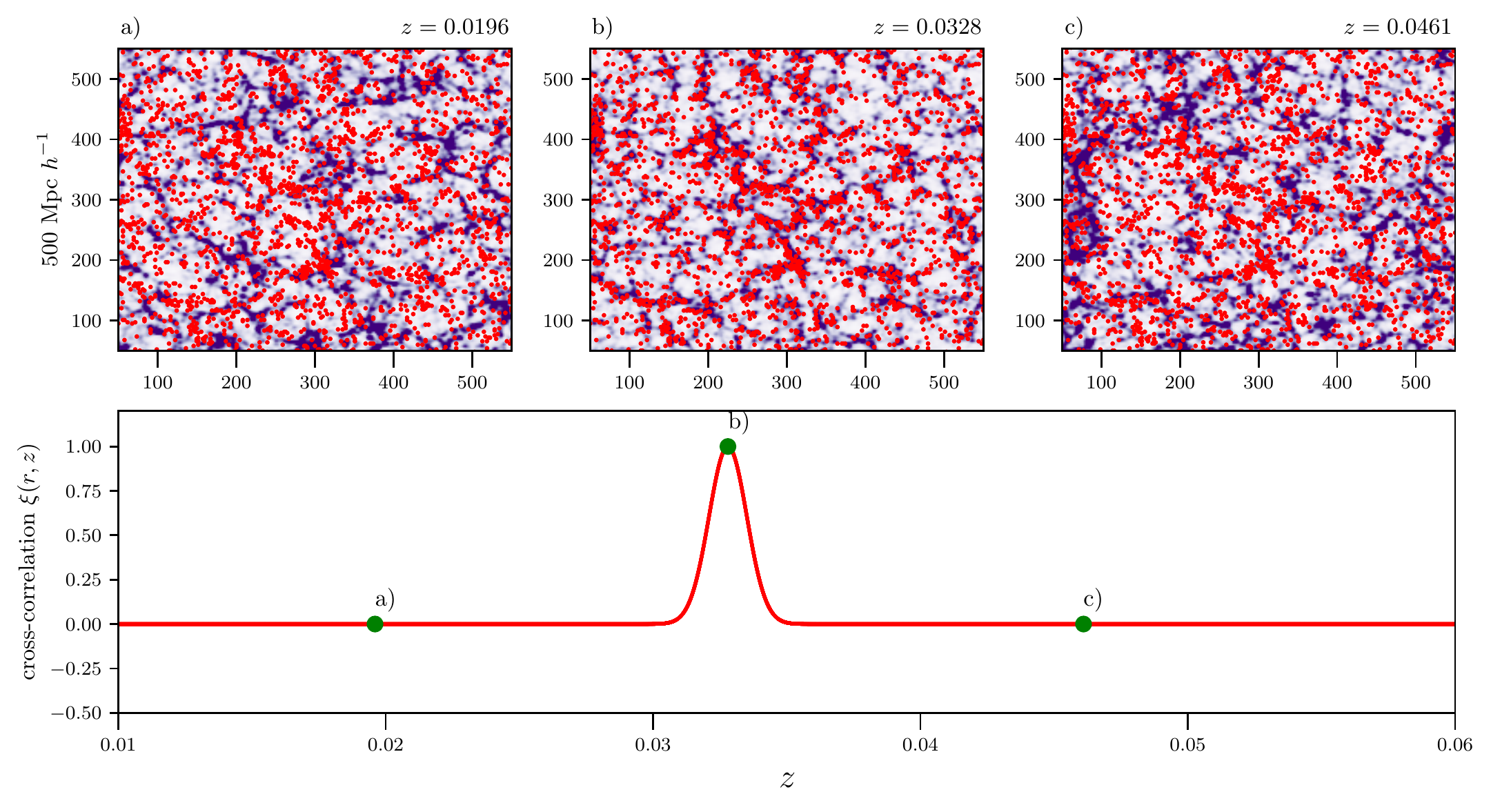}
\caption{The upper panel shows the clustering of simulated GW sources (red dots), for comparison with the simulated background matter distribution (shown in purple) at redshifts (a) $z= 0.0196$, (b) $z= 0.0328$, and (c) $z = 0.0461$, respectively. 
We show a $500\mpch\times500\mpch$ subsection of a cosmological simulation box.
The actual redshift of the simulated red points is $z = 0.0328$. The comparison of the red points with the purple background shows no correlation whatsoever for (a) and (c) whereas a strong correlation can be seen in (b) which represents the true redshift of the GW sources. The measured cross-correlation signal $\xi(r,z)$ can be represented as a Gaussian and  is shown in the bottom panel. The green dots correspond to the measured correlations for the cases (a), (b) and (c) respectively. As expected, the measured signal is essentially zero for the cases (a) and (c). For the case denoted by (b), the correlation signal is the strongest. The red curve is the expected correlation curve as a function of redshift. The number of gravitational wave sources have been exaggerated for visual clarity.
} 
\label{fig:corrLssNone}
\end{figure*}

In Fig.~\ref{fig:corrLssNone}, we demonstrate these concepts schematically. In each panel of the figure, the background grayscale is used to depict the projected matter density field from a slice from a cosmological simulation of dark matter with a width of $500 \mpch$ at three different distances. The red dots represent the positions of a large sample (for the purpose of illustration) of gravitational wave events in a given luminosity distance bin whose redshift distribution is the quantity one is seeking here.  
The cross-correlation function $\xi(r,z)$ shown in the lower panel estimates how the over-density field of the matter distribution and that of the gravitational wave events are correlated. In panel (b), the gravitational wave distribution is tracing the large scale structure of the matter distribution, while the distributions can be visually seen to be uncorrelated in panels (a) and (c). The resultant cross-correlation shown in the bottom panel shows a maximum at the redshift corresponding to the middle panel, while it is zero at the redshifts corresponding to the other two panels. In this way, by comparing the clustering of GW sources of an unknown redshift with the clustering of the background large scale structure (galaxy and galaxy clusters of known redshifts) at different redshifts, and studying the resultant correlation signal (the red curve in the bottom panel), the redshift of the GW sources can be inferred. Note that we do not assume that the GW events and the galaxies trace the matter distribution in an unbiased manner. In fact this is not crucial for this method to work as long as the galaxy distribution have redshifts which go beyond the redshift distribution of the GW events under consideration.

In practice, one will need to contend with the sky-localization error of the gravitational-wave sources, which will tend to weaken the cross-correlation signal -- an affect that will naturally occur in our simulations below. The magnification due to the weak gravitational lensing of the gravitational-wave sources by the structure in the foreground traced by the galaxies can also lead to correlations between the galaxies and the gravitational-wave source population \citep[see, e.g.,][]{Oguri:2016dgk, Congedo:2018wfn}. These residual low-redshift cross-correlations are less dominant than the cross-correlation arising from large scale structure~\citep{{Oguri:2016dgk}}. Here we have assumed that the galaxy catalogs used for the cross-correlation analysis extend beyond the GW event redshift distribution and that the number density of galaxies at the redshifts of the GW events is sufficient to measure the cross-correlation functions with a reasonable signal-to-noise ratio. In case the redshift distribution of the GW events extends much farther beyond the galaxy distribution, the weak lensing magnification induced cross-correlations between these GW events and the lower redshift galaxies, which is a separate cosmological probe, will dominate \citep[see e.g.][]{Congedo:2018wfn}. In that case, we would have to exclude GW events at large luminosity distances from our analysis. Alternatively, a spectroscopic galaxy survey campaign could be carried out to map the brightest galaxies at redshifts of interest (such a survey need not have the host galaxies of the GW events). In our analysis, since the assumption holds true, we can safely ignore this weak lensing effect for the purpose of this paper.


\section{Simulations}
\label{sec3}

We test the method described above by simulating a number of scenarios, starting from the  overtly optimistic case of a few thousand to the more realistic case of a few dozen BBH mergers detected by a network of gravitational wave observatories. The former large-number study enables us to check for systematics in our method as well as test the limits of its performance, especially when the true galaxy hosts of the gravitational-wave events are entirely absent from the galaxy catalog.

We use the Big MultiDark Planck (BigMDPL) cosmological N-body simulation for this purpose~\citep{Klypin:2014kpa}. The BigMDPL simulates a representative cubic volume of the Universe with a comoving size $L_{\rm box}=2.5~\gpch$ on each side and utilizes $3840^3$ particles with a mass resolution of $2.36\times10^{10}~\msunh$. The simulation uses a flat $\Lambda$CDM background cosmology, with the Hubble parameter $h = H_0/(100~\kmspMpc) = 0.678$, the matter density parameter $\Omega_{\rm m} = 0.307$, the amplitude of density fluctuations characterized by $\sigma_8 = 0.823$, and the power spectrum slope of initial density fluctuations, $n_s = 0.96$. These cosmological parameters are compatible with \textit{Planck} 2015 and 2018 results \citep{Ade:2015xua,Planck2018}. 

In order to create a mock observed galaxy catalog out of this simulation, we only use well resolved dark matter halos with masses above $M_{\rm th} = 10^{13.7}\msunh$. We assume that all galaxies are central galaxies and thus place each galaxy at the center of these dark matter halos. We place an observer at the centre of the box, and compute the sky positions of each of these mock galaxies. We also use the cosmological parameters of the simulation in order to convert the comoving distances of each of these galaxies from the observer to an 
observed cosmological redshift for these galaxies~\footnote{Since we consider the angular cross-correlation function, we do not account for the redshift space distortions.}. All cosmological calculations have been performed using the package \textsc{aum} publicly available at \url{https://github.com/surhudm/aum}. Given that we place the observer at the center of the box, the galaxies will be distributed out to a comoving distance $L_{\rm box}/2=1.25 \gpch$, which corresponds to a redshift range $z\in[0, 0.45]$.

As the particle database for the simulation was not publicly available, we randomly sample GW sources from dark matter halos with masses $M \in [10^{12.6},10^{12.9}]\msunh$ which are expected to be unbiased tracers of the matter distribution. The true luminosity distances for these sources are computed assuming the cosmological parameters of the simulation box and their comoving distances from the observer at the center of the box. Similarly the true sky locations of the GW sources are obtained in terms of right ascension (RA) and declination (dec) by projecting the sources on a unit sphere. For the simulation of the realistic GW sources, we choose only those sources whose true luminosity distances are within $900 \mpch$ from the observer\footnote{We find that beyond this distance range the event localization is quite poor for our choice of the simulation parameters.}. 

We inject gravitational wave sources at these sky positions. The distributions of the various injection parameters that are used for the simulation of these GW sources have been summarized in Table \ref{sourcepms}. The masses $m_1$ and $m_2$ of the two black holes have been chosen uniformly from the interval $10 M_\odot - 35 M_\odot$. This is the typical mass range observed so far in the previous LIGO-Virgo runs. The black hole spin parameter is denoted by $\chi_{1,2}$ and the magnitude is chosen to lie uniformly in the interval $(0,0.8)$. The spins can be aligned or anti-aligned. The angle between the two black hole spins is denoted by $\phi_{12}$. Also, $\phi_{jl}$ is the the angle between the orbital angular momentum and the total angular momentum of the system (orbital + spin). We simulate the time streams of data based on these injection parameters, the sky position, the luminosity distance, and   event times that are randomly chosen from a single sidereal day (in order to allow for a good sampling of detector orientations).

The simulated BBH signals are added to simulated Gaussian-noise strain data with Advanced LIGO Zero-Detuned-High-Power sensitivity~\citep{TheLIGOScientific:2014jea,aligo-zdhp} for the two LIGO detectors in the US and with Advanced Virgo sensitivity~\citep{TheVirgo:2014hva} for the Virgo detector in Europe. We only chose those events for the $H_0$ measurement where at least two of the detectors registered a signal-to-noise ratio (SNR) above a threshold value of 5 each, the third an SNR greater than 2.5, and combined SNR from all three detectors was found to be greater than 8. The posterior distributions of the corresponding BBH parameters were constructed directly from those time streams by employing the publicly available Bayesian inference code \textsc{bilby}~\citep{bilby}.  
\begin{table}
\centering
\begin{tabular}{ |p{2cm}|p{2cm}|p{2cm}|p{3cm}|p{3cm}|  }
 \hline
 \multicolumn{3}{|c|}{Injection Parameters} \\
 \hline
 Parameters & Distribution & Limits\\
 \hline
 $m_{1,2}$ & uniform & [10, 35] $M_\odot$ \\
 $\chi_{1,2}$ & uniform & [0, 0.8]   \\
 $\phi_{12}$ , $\phi_{jl}$  &  uniform & [0, 2$\pi$)  \\
 $\cos\theta_{1,2}$ , $\cos\iota$ & uniform & [-1, 1) \\
 $\psi$ , $\phi_{\rm c}$ & Fixed & 0 \\
 \hline
\end{tabular}
\caption{Injection parameters used for simulating realistic GW sky localizations using Advanced LIGO and Advanced Virgo detectors.
Here $\psi$ and $\phi_c$ are the polarization angle and the coalescence phase of the BBH signals.
The values of other parameters -- RA, Dec and $d_{\rm L}$ -- are taken from the Big MultiDark Planck simulation.}
\label{sourcepms}
\end{table}  

As we sample the GW sources from the matter distribution, we ensure that they are not located in the mock central galaxies or even within the same halos that host these galaxies in our mock catalogs. 
However, the GW events and the mock galaxies share the same underlying large scale structure and, therefore, are correlated with each other. 

As detailed in the next section, the measured BBH parameters are used to compute  the cross-correlation of the GW sources (binned in their inferred luminosity distance) with mock galaxies (of much more precisely known sky positions and redshifts). Following the method described in Sec.~\ref{sec2b}, the redshift distribution of the GW sources is deduced using  that cross-correlation. We take equal luminosity-distance bins of fixed width $\Delta d_{\rm L} = 200$ Mpc. Sources that are very close ($d_{\rm L} \leq 200$ Mpc) are excluded from the analysis since such sources are rare and do not statistically contribute. We consider only those sources that have an observed $d_{\rm L}$ (taken to be the median of the $d_{\rm L}$ posterior) 
in the range $[200-1400]$ Mpc and an overall three-detector network SNR $>8$. The total number of such sources in our first (large) simulation that satisfied these criteria turned out to be $\sim5100$. The mock galaxies are also binned in 20 equal redshift bins of $\Delta z = 0.015$. The binning for the luminosity distance is chosen such that there are enough sources per bin so as to result in a cross-correlation with a well-defined peak. The bins in redshift distributions are chosen such that the peak of the cross-correlation function is covered with multiple points. Increasing the luminosity distance bin width increases the significance of the measured cross-correlation but widens the redshift distribution. Increasing the redshift bin width will reduce the cross-correlation signal as the cross-correlation signal gets integrated over a longer line-of-sight interval. In order to compute the cross-correlation signal, we use the location that corresponds to the {\em  maximum} value of the posterior distribution, as the sky position of each gravitational wave source. While unrealistic in the current generation of detectors, this large simulation is nevertheless useful in checking if our method has any inherent parameter estimation bias.

\section{Cross-correlation analysis and results}
\label{sec4}

For every luminosity distance bin, we compute the cross-correlation of gravitational wave events in that bin with galaxies in each of the 20 redshift bins. This results in 20 values of the cross correlation for each luminosity-distance bin. On average, one expects the cross correlation to peak for the redshift bin that most accurately represents the chosen luminosity-distance bin.

We use the simple Peebles-Davis estimator for the two-point (angular) cross-correlation between the GW sources with unknown redshifts and galaxies with known redshifts\footnote{In real data, with masked out areas and irregular galaxy catalog shapes, the Landy-Szalay estimator would be more appropriate.}, such that
\begin{equation}
    w(\leq \theta_{\rm max}) = \frac{n_{\rm D_1 D_2}(\leq \theta_{\rm max})}{n_{\rm R_1 R_2}(\leq \theta_{\rm max})}-1 \,,
    \label{correstimator}
\end{equation}
where $n_{\rm D_1 D_2}(\leq \theta_{\rm max})$ is the number of galaxy-GW source pairs with an angular separation of $\theta_{\rm max}$ or less, as seen by the observer, while $n_{\rm R_1 R_2}(\leq \theta_{\rm max})$ is the number of galaxy-GW source pairs expected with angular separations less than $\theta_{\rm max}$ if the galaxies and GW sources were randomly distributed on the sky. While determining what the correlation angle $\theta_{\rm max}$ should be, care must be taken so that $\theta_{\rm max}$ is not too small compared to the average $ 1 \sigma$ error region for the sky localization of the GW sources. Increasing $\theta_{\rm max}$ significantly also washes out the correlation signal since any existing correlation would then be averaged over a larger region. Thus, an optimum value of $\theta_{\rm max}$ has to be chosen such that the cross-correlation is maximum for that value of $\theta_{\rm max}$. 
In our case, we determine the value of $\theta_{\rm max}$ by finding the $\theta_{\rm max}$ that gives a relatively large signal to noise ratio for the angular cross-correlation signal (see Fig. \ref{snrtheta} in the Appendix~\ref{sec:app1}). For our simulated data set, $\theta_{\rm max}$ is found to be $0.03$ rad, which is $\sim 1.7$ deg, and corresponds to a projected comoving distance of $\sim 3$ Mpc at $z = 0.02$ and $\sim 36$ Mpc at $z= 0.3$ (well beyond the virial radii of the halo). This value of $\theta_{\rm max}$ was kept fixed over the entire $d_{\rm L}$ range \citep[c.f.][]{Newman:2008mb}, who use a fixed value of physical distance $r_{\rm max}$ within which they integrate the angular correlation function. For every luminosity-distance bin of GW events, we thus measure a single value for its cross-correlation strength corresponding to each redshift bin of galaxies.

\begin{figure*}
\centering
 \includegraphics[width = \textwidth, height=0.8\textwidth]{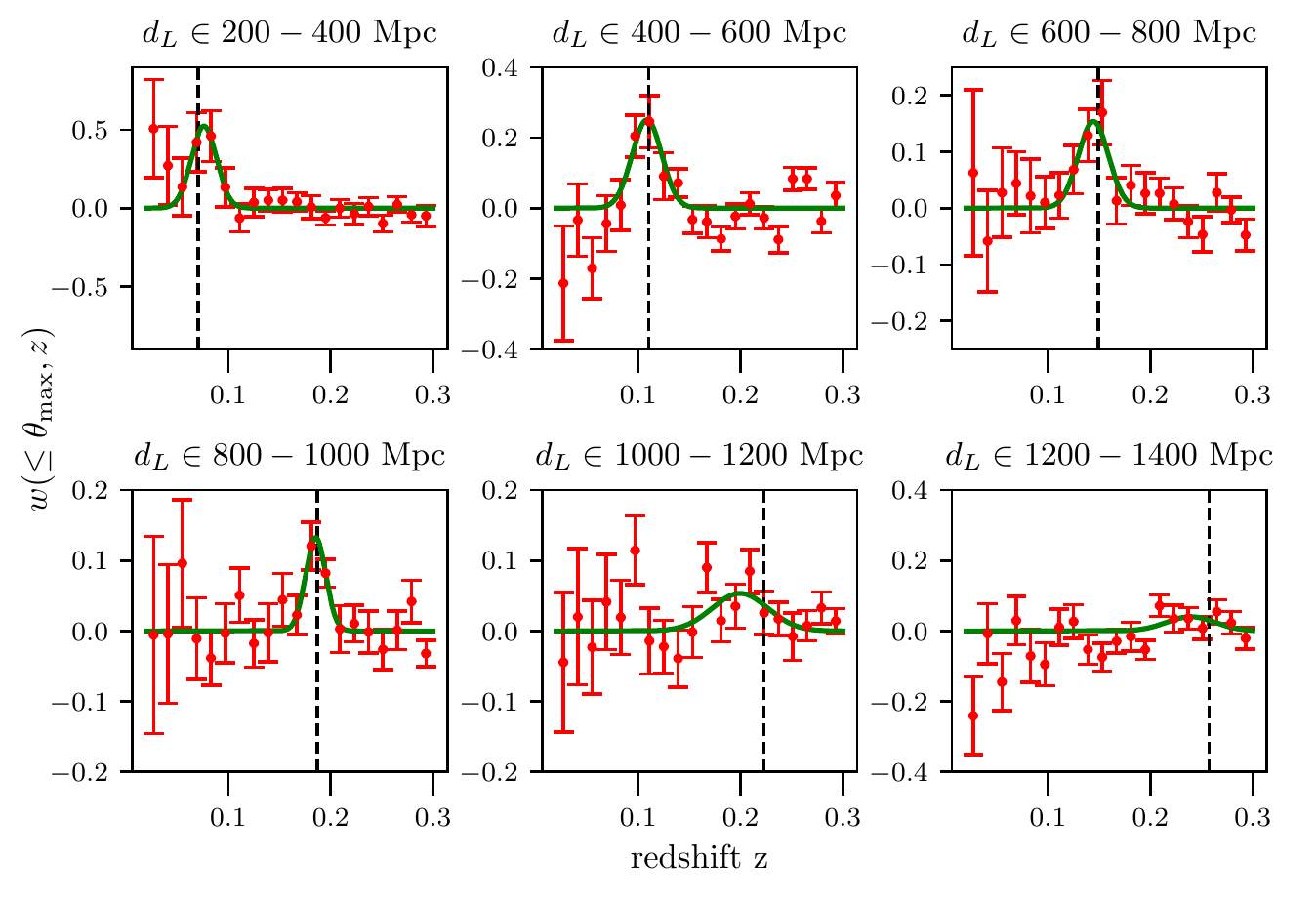}
\caption{The angular cross-correlation $w(\leq \theta_{\rm max}, z)$  for six different luminosity distance bins for a realistic simulation with 5100 simulated sources scattered upto $\sim 1400$Mpc in luminosity distance. For each bin, there is an associated true redshift distribution of the GW sources in that bin. This is captured by the peak of the angular cross-correlations of the sources with the galaxy distribution. The red points are the measured cross-correlations with error bars, while the green curve is a Gaussian fit to the measured distribution. The black dashed curve shows the {\textit true} average redshift of the sources in a given bin. The statistical errors are larger for redshift bins with relatively less number of halos.} 
\label{corr1}
\end{figure*}

In Fig.~\ref{corr1}, we show the angular cross-correlation $w(\leq \theta_{\rm max},z)$ as a function of the redshift $z$ for 6 different $d_{\rm L}$ bins. The $x$-axis corresponds to the 20 redshift bins obtained from the galaxy distribution. Each box in Fig.~\ref{corr1} corresponds to a particular $d_{\rm L}$ bin (with range shown on the $x$-axis in each box). The red points are the Jackknife mean of $w(\leq \theta_{\rm max})$, and the error bars are the corresponding Jackknife standard deviation estimations $\sigma_w$ (the Jackknife method is used in order to obtain the bias of the estimator $w(\leq\theta_{\rm max})$. It is one of the widely used resampling techniques apart from Bootstrap \citep{Jackknife}). The black dashed line is the {\it true} average redshift of the GW sources in a given $d_{\rm L}$ bin that is calculated using the standard values of the $\Lambda$CDM model parameters and taking $H_0= 70 \kmspMpc$. 

{\bf In principle, we could construct a proper forward model in order to predict each of these cross-correlation functions given the value of the Hubble constant \citep{Oguri:2016dgk}. However, we note that we can construct a Hubble diagram, if we infer the mean redshift of the GW event distribution and its error.  Given that our cross-correlation function is expected to correspond to a distribution with a mean and variance, we choose the distribution to be a Gaussian in order to minimize any prior information \citep{HoggLang}.} Therefore, we fit the measured cross-correlation in each panel with
\begin{equation}
    w(\leq \theta_{ \rm max}, z, z') \propto \exp\left[-\frac{(z-z')^2}{2\sigma_{z}^2}\right]
    \label{gaussian}
\end{equation}
in order to obtain the mean redshift  $z'$ corresponding to each $d_{\rm L}$ bin and $\sigma_{z}$, which gives the spread in redshift. We are interested in the error on the mean, $\sigma_{z'}$ on $z'$ from the covariance matrix of the estimated parameters of the Gaussian.

The green solid curve in Figure~\ref{corr1} is a Gaussian fit to the red points assuming the form given in Eq.~\eqref{gaussian}. The curve is fitted using standard non-linear least-square method with each data point assigned a weight of $1/\sigma_w^2$ corresponding to the Jackknife error as shown in the $y$-axis of Fig.~\ref{corr1}. The peak of this Gaussian corresponds to the {\it true} redshift of the GW sources in a given bin. 

\begin{figure}[h!]
\centering
 \includegraphics[width = \linewidth]{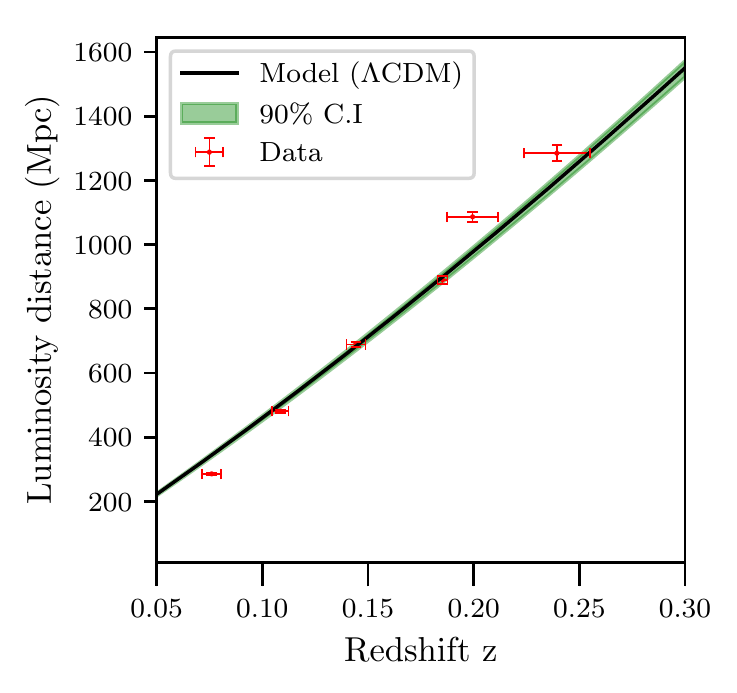}
\caption[Optional caption for list of figures]{Luminosity distance (Mpc) plotted against the inferred redshifts for a realistic simulation of 5100 sources with sources spread up to $d_{\rm L} \sim 1400$ Mpc in luminosity distance. The error bars in redshift have been obtained by assuming a Gaussian distribution as shown in Fig.~\ref{corr1}, while the error bars in the luminosity distance have been obtained from the posterior distribution of $d_{\rm L}$. The black solid line is the line corresponding to the flat $\Lambda$CDM model with $H_0 = 70 \kmspMpc$. The shaded region represents $90\%$ confidence interval of the $H_0$ posterior around the median. For ease of visualization of the errors in the luminosity distance, we plot the error bars along the $y$-direction to be 3 times the actual values.
} 
\label{hubble1}
\end{figure}

For each luminosity distance bin, we calculate a corresponding $z'$, and its associated error for the redshift distribution, $\sigma_{z'}$. We assign an error-weighted average luminosity distance to the binned sources where the errors on the luminosity distance comes directly from the parameter estimation results from the simulated gravitational wave events. This resultant luminosity distance {\it vs} redshift relation is shown in Fig.~\ref{hubble1}. The red points along with the error bars in redshift and $d_{\rm L}$ directions are obtained from our cross-correlation analysis. Each red point corresponds to a particular $d_{\rm L}$ bin of Fig. \ref{corr1}. The black solid line is the $\Lambda$CDM model predicted line with $H_0=70$ $\kmspMpc$, which is the true value of $H_0$ used in the simulation. The vertical errors on the luminosity distance are fairly small given the large number of sources in each bin in this plot. For ease of visualization of the errors in the luminosity distance, we plot the error bars along the $y$-direction to be 3 times the actual values.

In a flat $\Lambda$CDM cosmology and the redshift range of our interest, the luminosity distance is related to the redshift as :
\begin{equation}
    d_{\rm L}(z) = \frac{c(1+z)}{H_0}\int_0^z \frac{dz'}{\sqrt{1-\Omega_{\rm m}+\Omega_{\rm m} (1+z')^3}}\,.
\end{equation}

We now perform a Bayesian analysis to find the posterior of the Hubble parameter $H_0$ with the help of a Monte Carlo Markov chain sampler. We discuss the details of the posterior distribution in the following subsection. Given that we exclusively focus on the Hubble parameter in this paper, we fix the value of $\Omega_{\rm m}$ to correspond to the true value of the matter density parameter of the simulation.

\subsection{Bayesian estimates}
\label{sec4a}

Following Bayes' theorem, the posterior probability density of $H_0$ for a given set of inferred values of $D_{\rm gw} \equiv \{d_{{\rm L},i}, z_i\}$ is given by,

\begin{equation}
    P(H_0|D_{\rm gw}) = \frac{P(H_0)P(D_{\rm gw}| H_0)}{\int_{H_0} P(H_0) P(D_{\rm gw}| H_0 )\,dH_0}
    \label{likeli}
\end{equation}
where $P(H_0)$ is the prior probability density distribution of $H_0$. The probability $P(D_{\rm gw}| H_0)$ is the likelihood of observing the set of $\{d_{{\rm L},i}, z_i\}$, given a value of $H_0$. The term in the denominator of Eq. \eqref{likeli}, is the evidence. We consider a flat uniform prior on $H_0$ between $50 \kmspMpc$ and $90 \kmspMpc$. The posterior probability density function(\textit{pdf}) is therefore given by
\begin{equation}
    P(H_0|D_{\rm gw}) \propto P(D_{\rm gw}| H_0)\,.
\end{equation}

If $\mathcal{L}_i = P(D_{{\rm gw},i}| H_0)$ denotes the likelihood of each individual data point $D_{{\rm gw},i}$, then the overall likelihood $P(D_{\rm gw}| H_0)$ would be given by,
\begin{equation}
\mathcal{L} = P(D_{{\rm gw}}| H_0) = \prod_i P(D_{{\rm gw},i}| H_0)
\label{likelihood}
\end{equation}
For the computation of the likelihood, we account for errors in both the $x$ and $y$-directions \citep{HoggLang}. Thus the computation of the likelihood $P(D_{{\rm gw},i}| H_0)$ involves an integral over the entire redshift direction and is given by
\begin{eqnarray}
  \mathcal{L}_i = P(D_{{\rm gw},i}| H_0) &\propto& \int \exp\left[-\frac{[d_{{\rm L},i}- d_{\rm L}(z,H_0)]^2}{2\sigma_{d_{{\rm L},i}}^2}\right] \nonumber \\
  &\times& \exp\left[-\frac{(z_i - z)^2}{2\sigma_{z,i}^2}\right] dz
\end{eqnarray}
where $d_{\rm L}(z,H_0)$ is the expected value of $d_{\rm L}$ at a redshift $z$ for a given $H_0$ and $d_{{\rm L},i}$ is the measured value denoted as the red points in Fig. \ref{hubble1}. The redshift $z$ denotes the possible true average value of the redshift distribution of our GW sources, whereas $z_i$ is the redshift inferred from the cross-correlations. The quantities $\sigma_{d_{{\rm L},i}}$ and $\sigma_{z,i}$ are the standard deviations in the respective measurements. 

The final posterior probability distribution of $H_0$ is obtained by using Eq. \eqref{likelihood} in Eq. \eqref{likeli}. We sample from this posterior distribution by using an MCMC sampler initiated randomly within the prior range $[50,90] \kmspMpc$. For this purpose, we use the python based MCMC package \textsc{emcee} \citep{emcee}. The resulting posterior of $H_0$ is shown below in Fig. \ref{hpost2}. 

\begin{figure}[h!]
\centering
\includegraphics[width = \linewidth]{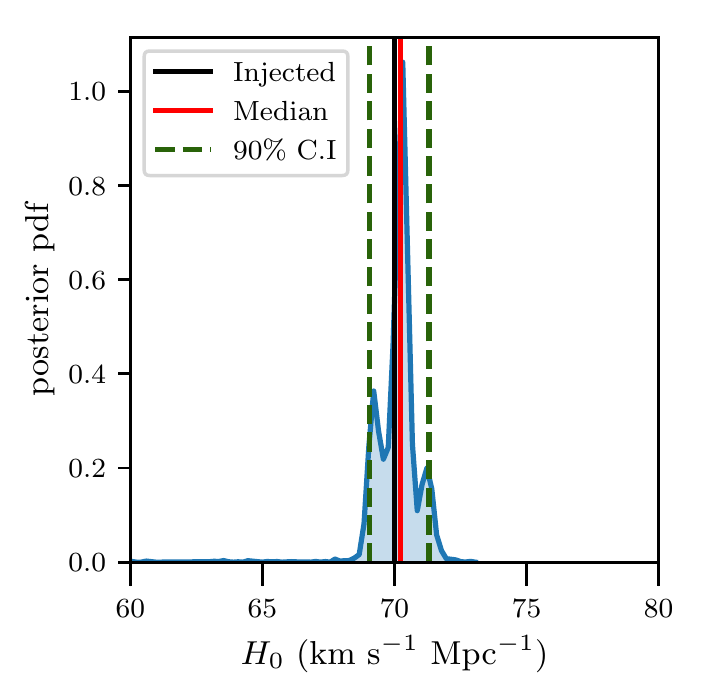}
\caption[Optional caption for list of figures]{The posterior probability density for $H_0$ corresponding to fig. \ref{hubble1} for a realistic simulation of 5100 GW sources. The prior on $H_0$ has been taken as uniform in the interval $[50,90] \kmspMpc$. The black line is the injected value $H_0 = 70 \kmspMpc$ used for the source simulations. The red line is the median of the posterior. The green dashed lines mark the boundary of the $90\%$ confidence interval.} 
\label{hpost2}
\end{figure}

In Fig. \ref{hpost2}, we show the posterior \textit{pdf} for the Hubble parameter $H_0$ considering sources upto a redshift of $z \leq 0.3$. The injected value used in simulating the sources is $H_0 = 70 \kmspMpc$. The Bayesian analysis, as described above, gives a posterior distribution of $H_0$ with the constraints $H_0 = 70.22^{+1.09}_{-1.18} \kmspMpc$, which defines the $90\%$ confidence interval around the median value $70.22 \kmspMpc$. In Fig. \ref{hpost2}, the median value  $H_0 =70.22 \kmspMpc$ has been indicated with a red solid line. The region between the green dashed lines denotes the $90\%$ confidence interval around the median $H_0$. This interval is depicted as a shaded region in Fig. \ref{hubble1}. The injected value $H_0=70 \kmspMpc$ has been denoted by the black solid line. In this case, the posterior of $H_0$ is almost symmetric and the median differs only slightly from the injected value. The shaded region in the plot is the posterior probability distribution of $H_0$. \\

\begin{figure}[h!]
\centering
 \includegraphics[width = \linewidth]{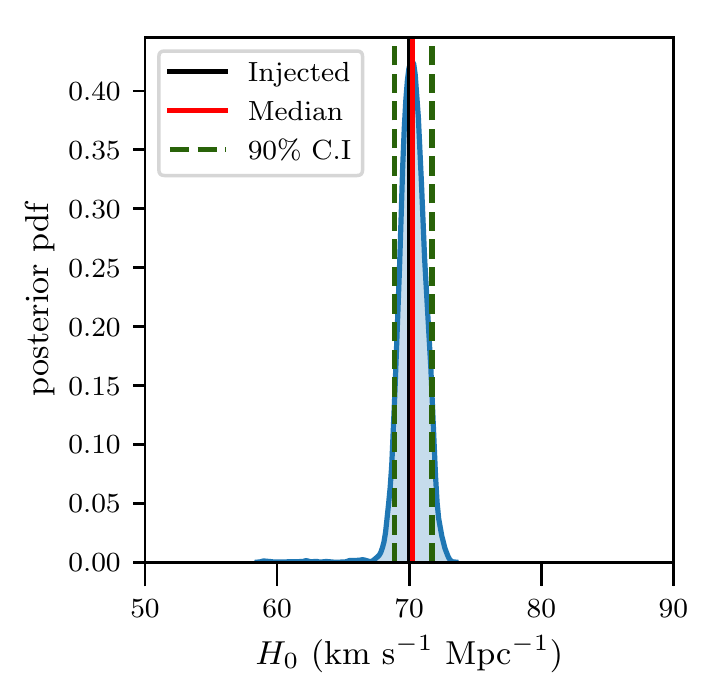}
\caption[Optional caption for list of figures]{The posterior probability density of $H_0$ for a realistic simulation of 500 sources. The prior on $H_0$ has been taken as uniform in the interval $[50,90] \kmspMpc$. The black line is the injected value $H_0 = 70 \kmspMpc$ used for the source simulations. The red line is the median of the posterior. The green dashed lines mark the boundary of the $90\%$ confidence interval.} 
\label{hpost3}
\end{figure}

\begin{figure}[h!]
\centering
 \includegraphics[width = \linewidth]{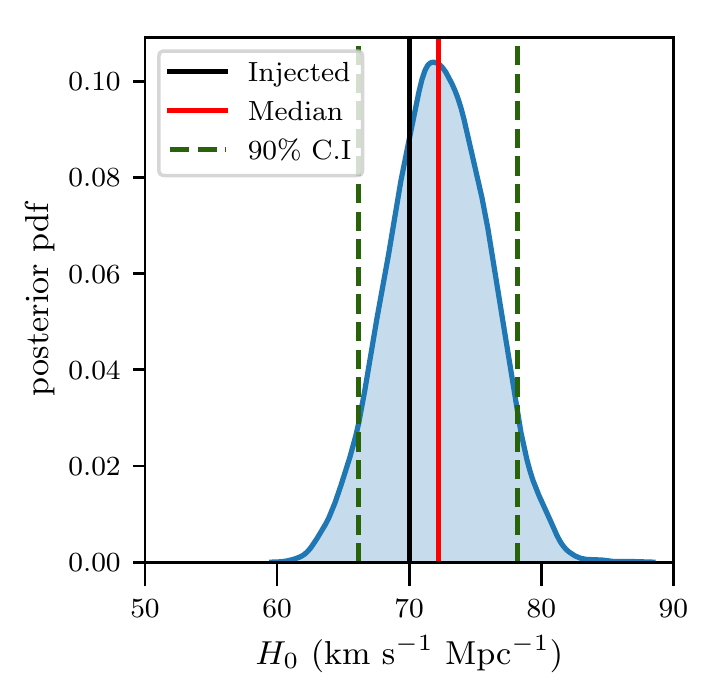}
\caption[Optional caption for list of figures]{The posterior probability density of $H_0$ for a realistic simulation of 50 events. The prior on $H_0$ has been taken as uniform in the interval $[50,90] \kmspMpc$. The black line is the injected value $H_0 = 70 \kmspMpc$ used for the source simulations. The red line is the median of the posterior. The green dashed lines mark the boundary of the $90\%$ confidence interval.} 
\label{hpost4}
\end{figure}

We repeat our analysis for even less number of randomly selected GW sources to test the dependence of the constraints on the number of GW sources that we use for cross-correlations. For this purpose, we choose $\sim 500$ and $\sim 50$ nearby GW events out of the previously generated $5100$ simulations, such that the luminosity distance of these events $\in [200, 900]$Mpc. Using the same correlation technique, we perform an MCMC on these samples assuming a uniform prior of $H_0$ in the range $[50,90] \kmspMpc$. Figures \ref{hpost3} and \ref{hpost4} show the posterior distributions of $H_0$ and the corresponding $90\%$ confidence intervals for $\sim 500$ and $\sim 50$ observations, respectively~\footnote{These numbers are consistent the sample sizes one expects with one year to several years of observations with Advanced LIGO and  Advanced Virgo~\citep{LIGOScientific:2018mvr}}. The red line is the median of the posterior distribution while the green dashed lines are the boundary of the $90\%$ confidence interval around the median. The black vertical line is the injected value $H_0=70 \kmspMpc$. The constraints coming from $500$ sources are $H_0 = 70.26^{+1.47}_{-1.40} \kmspMpc$ and that from $50$ sources are $H_0 = 72.24^{+5.98}_{-6.05} \kmspMpc$. As expected, the posterior broadens when the number of GW events is less.\\

In Table \ref{resulttable}, we summarize the constraints on $H_0$ as obtained using the cross-correlation method.

\begin{table*}
\centering
\begin{tabular}{ |p{3cm}|p{3cm}|p{3cm}|p{3cm}|  }
\hline
 \multicolumn{4}{|c|}{Constraints on $H_0$} \\
 \hline
 No. of GW events & Max $d_{\rm L}$ (Mpc) & Injected $H_0$ (km s$^{-1}$ Mpc$^{-1}$)  & Constraints on $H_0$ (km s$^{-1}$ Mpc$^{-1}$)\\
 \hline
 $5100$ & 1400 & 70 & $70.22^{+1.09}_{-1.18}$   \\
 $500$ & 900 & 70 & $70.26^{+1.47}_{-1.40}$  \\
 $50$ & 900 & 70 & $72.24^{+5.98}_{-6.05}$ \\
 \hline
\end{tabular}
\caption{Constraints on $H_0$ using the cross-correlation method. The table summarizes the results obtained for 
5100,
500, and 50 sources, all simulated with realistic posteriors obtained from full parameter estimation runs using the \textsc{bilby} code.
The measurement precision improves as the number of detected GW events increases. The error bars indicated define the $90\%$ confidence intervals around the median as quoted in each case.}
\label{resulttable}
\end{table*}

\section{Discussions and Summary}
\label{discussions}

In this paper, we have demonstrated how BBH sources can be used to constrain the value of the Hubble constant $H_0$, even when the host galaxies of these sources are entirely absent in the galaxy catalogs. The localized positions of the BBH events obtained by a network of gravitational wave detectors can be cross-correlated with existing galaxy catalogs in order to infer the redshift distribution of the BBH events. We combine the inferred redshift information with the measurements of the luminosity distances from the GW detectors in order to infer the value of the Hubble constant. 

The current state of the art 
cross-correlation 
methods employed to analyze BBH data~\citep{Gray:2019ksv, Abbott:2019yzh} have obtained respectable constraints on the Hubble constant. However, the formalism in these studies assumes the GW events to be randomly distributed with no regard to the clustering of these sources or to their cross-correlation with galaxies on large scales. Therefore, these studies have to rely on the presence of individual host galaxies in the catalog, which leads to an inherent dependence of the resultant constraints on implicit priors regarding the relation of the gravitational wave events and properties of galaxies in the catalog. We have shown that we can do away with these priors by using the information present in large scale structure. Especially at redshifts corresponding to BBH events, galaxy catalogs are expected to be incomplete. Thus, there is a significant gain to be had even if the host galaxy is not present in the catalog. The bright end of the galaxy population can be sampled in order to infer the redshift distribution of the gravitational wave sources by relying on the cross-correlation technique.

In our work, we have simulated different numbers of GW events drawn from the matter distribution in a large cosmological simulation. As we start from realistic simulations of the time series data for these GW events, our simulations have realistic uncertainties on the sky localization of these events as well as realistic cross-correlation of these events with large scale structure. We have measured the cross-correlation between these simulated GW sources and a mock sample of galaxies at different redshifts and showed that we can infer the redshift distribution of the GW sources without relying on the presence or absence of their true hosts. We show that posterior distribution of $H_0$ can be obtained with an accuracy of less than 10\% even with a sample of 50 events.

For simplicity of the analysis, we have considered a somewhat idealistic condition where the GW sources are unbiased tracers of the underlying matter distribution. In reality, the GW events as well as the galaxies used to trace the large scale structure will have a redshift dependent bias. Therefore while working with real data, these effects will have to be parameterized and marginalized over in order to obtain cosmological constraints. In future, we will implement a full Bayesian approach that models the measurement of the clustering of GW sources, galaxies, and their cross-correlations in a self-consistent framework. {\bf This will remove the need for characterizing the mean of the redshift distribution of the gravitational wave sources as a Gaussian distribution}. Such an analysis can also naturally incorporate the effects of the weak lensing of gravitational wave sources due to the intervening large scale structure, which results in a non-trivial cross-correlation at redshifts lower than that of the GW events.

With a large pool of GW data expected in the next decade or so, this method can prove to be very powerful in determining the various important cosmological parameters. Simulations of five-year data using Advanced LIGO-Virgo network suggest less than $5\%$ accuracy in the measurement of $H(z)$, even at a redshift of $0.8$ could be achieved \citep{Farr:2019twy}. \citet{Chen:2017rfc} shows that the constraints on the Hubble parameter might improve to $1\%$ within a decade or so. As we go deeper in redshift and higher in precision, with more number of sources and less uncertainties, the constraints are expected to be reduced even less, to a sub-percent level.

\section*{Acknowledgments}

We thank Chris Messenger, Remya Nair, Masamune Oguri, Aseem Paranjape and Masahiro Takada for helpful discussions. Thanks are also due to Maya Fishbach for carefully reading the manuscript and making useful suggestions. We thank the organizers and participants of the Fourth Physics and Astrophysics at the eXtreme (PAX-IV) workshop in IUCAA where the initial discussions on this work were seeded. We also thank the anonymous referee for a careful reading of the manuscript and useful comments. The CosmoSim database used in this paper is a service by the Leibniz-Institute for Astrophysics Potsdam (AIP). The MultiDark database was developed in cooperation with the Spanish MultiDark Consolider Project CSD2009-00064. The computing for this project was supported by the Pegasus cluster at IUCAA. This work was also supported in part by Tata Trusts.

\appendix

\section{Hubble diagrams for the simulated GW events}
\label{sec:app1}
\begin{figure}[h!]
\centering
  \includegraphics[width = \linewidth]{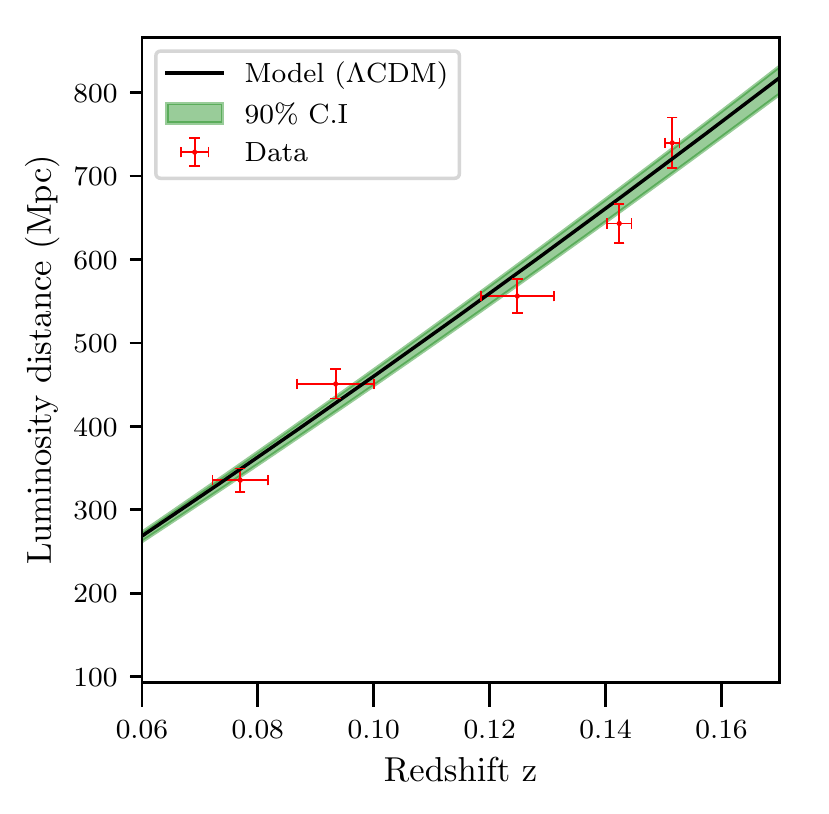}
  \includegraphics[width = \linewidth]{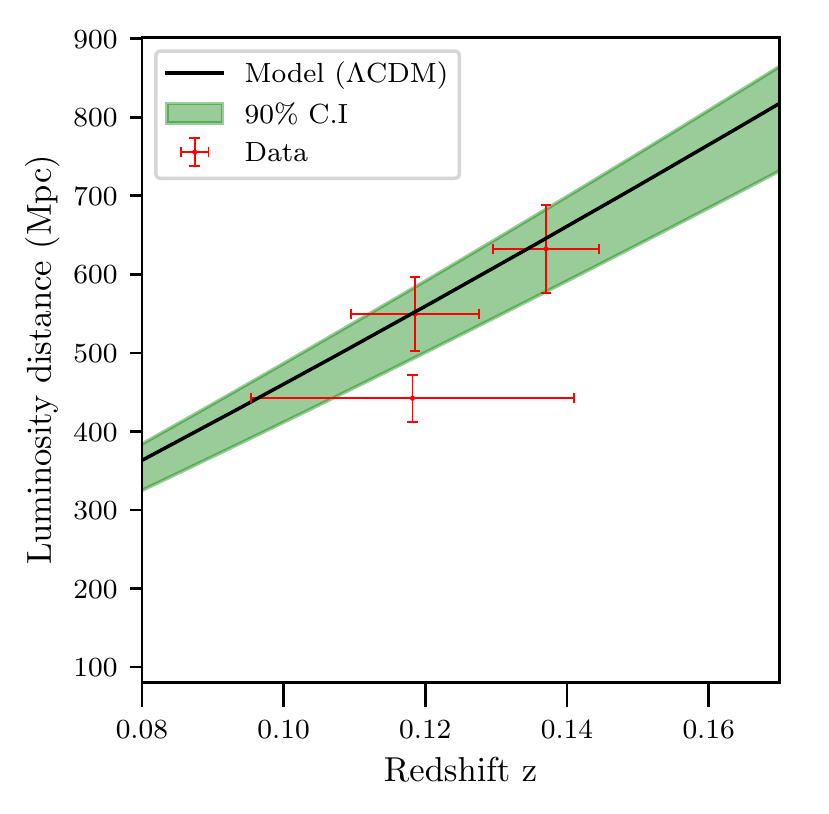}
\caption{Hubble diagrams for smaller subsamples of detected events from our simulated GW dataset. The top panel shows the plot of $d_{\rm L}$ {\it vs} redshift $z$ for 500 sources. The red points are the data. The inferred error in redshift is larger in this case than in Fig.~\ref{hubble1} due to smaller number of events considered. The black line corresponds to the standard $\Lambda$CDM model with $H_0 = 70 ~ \kmspMpc$ . The shaded region shows $90\%$ confidence interval of the $H_0$ posterior around the median. The corresponding posterior probability distribution of $H_0$ as obtained using the red data points is shown in Fig. \ref{hpost3}. The bottom panel is the same for a subsample of 50 GW events. The error bars are thus larger in this case. The resulting posterior is plotted in Fig. \ref{hpost4}. }
\label{fig:hubble50and500}
\end{figure}

In Sec.~\ref{sec4a}, we deduced the posterior probability of $H_0$ and the concomitant constraints from 500 and 50 simulated GW events in a three-detector LIGO-Virgo network with SNRs above 8. With the Advanced LIGO and Virgo design sensitivities, these projected measurements will be attainable within a matter of years. In Fig.~\ref{fig:hubble50and500}, we show the Hubble diagrams ($d_{\rm L}$ vs $z$) for these 500 and 50 simulated events that correspond to the posteriors presented in Figs.~\ref{hpost3} and \ref{hpost4}, respectively.

\begin{figure}[h!]
\centering
  \includegraphics[width = \linewidth]{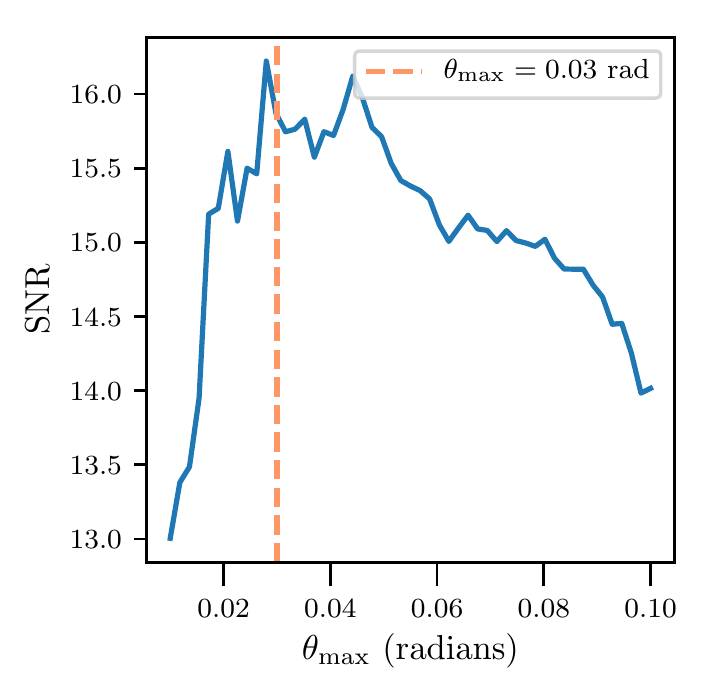}
\caption{Variation of the cross-correlation signal-to-noise ratio (SNR) for different values of $\theta_{\rm max}$. The SNR decreases for larger $\theta_{\rm max}$ due to the averaging out of the correlation signal at large scales. At the lower end of $\theta_{\rm max}$ (when $\theta_{\rm max}$ is smaller than the average angular localization scale of the GW sources), the SNR for the correlation measurements reduces sharply since there is a higher probability for the true source to lie outside the angular scale $\theta_{\rm max}$. The orange dashed line shows the value of $\theta_{\rm max}$ chosen for our analysis ($\theta_{\rm max} = 0.03$ radians) where the SNR value is optimum.}
\label{snrtheta}
\end{figure}

In Fig. \ref{snrtheta}, we show the variation of cross-correlation signal-to-noise ratio for different values of $\theta_{\rm max}$.

\section{Cross-correlation analysis with Gaussian sky localization area}
\label{sec:app2}

The $5100$ sources simulated in  Sec.~\ref{sec3} have realistic source distributions. To check our results obtained from realistic simulations, here we alternatively pursue a set of idealized measurements, namely, by using Gaussian distributions for the BBH sky localizations and luminosity distances. For the latter parameter, we set the measured value equal to the true (injected) one, and use a $1\sigma$ spread of  $10\%$ about it. The construction of the idealized sky localizations is detailed below.

For simulating mock galaxies, we use halos catalogued from the MultiDark Planck 2 (MDPL2) cosmological N-body simulation with a halo-mass threshold of $M_{th} = 10^{13}~ M_{\odot} h^{-1}$. The MDPL2 simulates a comoving volume of the universe corresponding to a comoving size $\sim 1 ~ \gpch$. The simulation is performed using $3840^3$ particles with a mass resolution $1.5 \times 10^9 ~ \msunh$ assuming a flat $\Lambda$CDM cosmology. The important paramater details can be found at \url{https://www.cosmosim.org/}. 

As discussed in Sec.~\ref{sec3}, we compute everything with respect to an observer at the center of the simulation box. We thus create a mock galaxy catalog with full sky coverage out to a distance of $500 ~\mpch$. The sky locations are computed by projecting the 3D locations of the galaxies on to a 2D unit sphere. We compute the cosmological redshifts of the galaxies from their comoving distances using the same cosmology used for the simulation. Relative to the center of the box, the redshifts of these galaxies  lie in the range $[0,0.18]$. We simulate a mock GW source catalog by randomly sorting $5000$ dark matter particles from the same simulation box within a spherical volume of radius $500~ \mpch$. The GW sources thus chosen are unbiased tracers of the matter distribution in the simulation. The observed luminosity distances for these sources are computed from their comoving distances away from the observer. For true sky locations, we use the same method as in the case of mock galaxies. For the sky localization error, we perturb the true locations of the simulated GW sources with a two-dimensional Gaussian distribution, with a spread equal to $0.03$~radians at $500~\mpch$, and the standard scaling with increasing distance.

\begin{figure*}
\centering
 \includegraphics[width = \textwidth, height=0.8\textwidth]{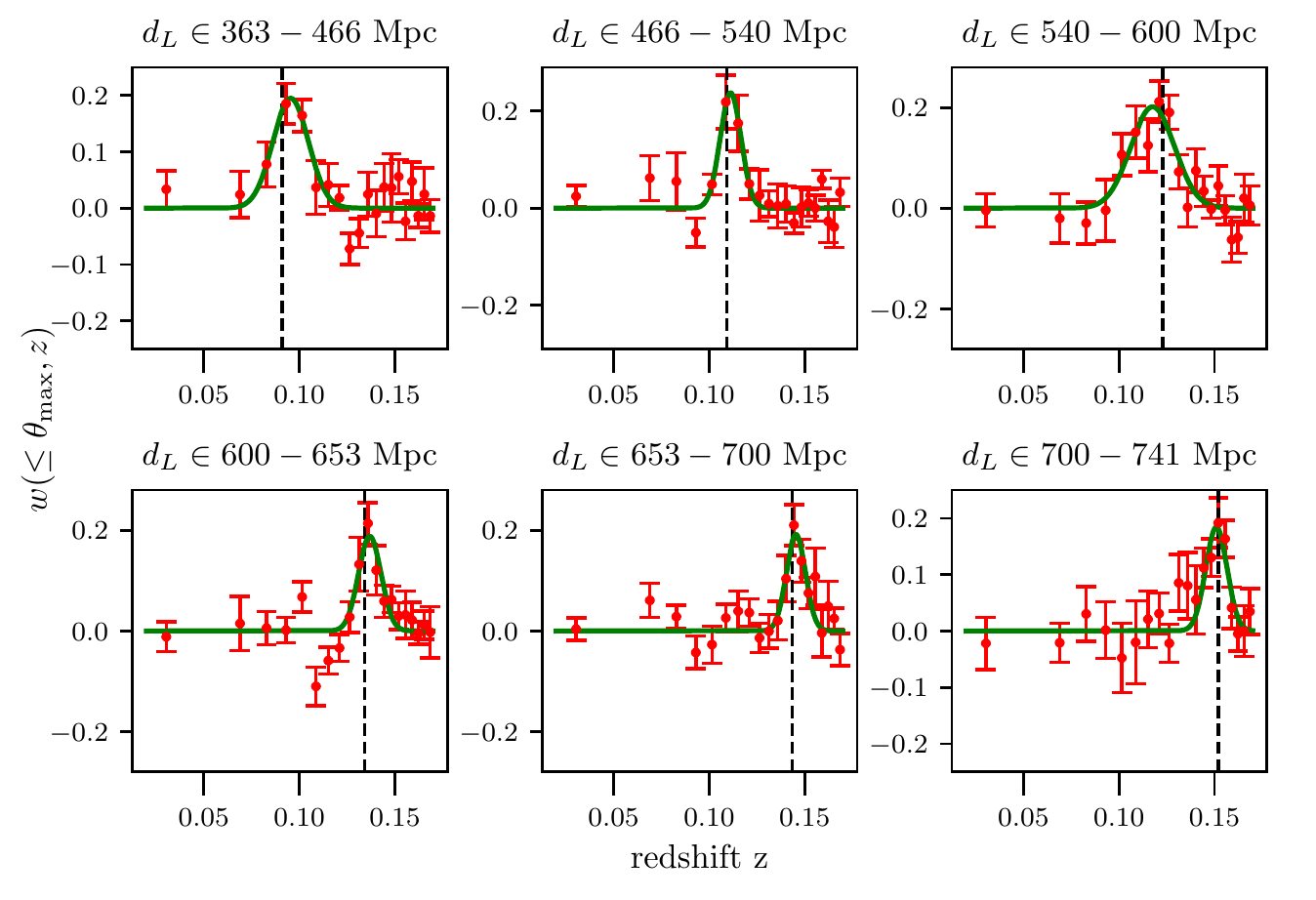}
\caption[Optional caption for list of figures]{The angular cross-correlation $w(\leq \theta_{\rm max}, z)$  for six different luminosity distance bins, just as in Fig.~\ref{corr1}, but now for the 5000 simulated BBHs with Gaussian sky and distance errors, as described in Appendix~\ref{sec:app2}.}
\label{corr}
\end{figure*}

The GW events are binned into six different luminosity distance bins such that each bin corresponds to the same comoving volume with the expectation of preserving similar signal-to-noise per bin. We bin the mock galaxy sample in 20 different redshift bins in a similar manner. 
The cross-correlation analysis can now be performed using the technique described in sec \ref{sec4} for each of the 6 luminosity distance bins. The corresponding redshift distributions recovered are shown in Fig.~\ref{corr}.

\begin{figure}[h!]
\centering
\includegraphics[width = \linewidth]{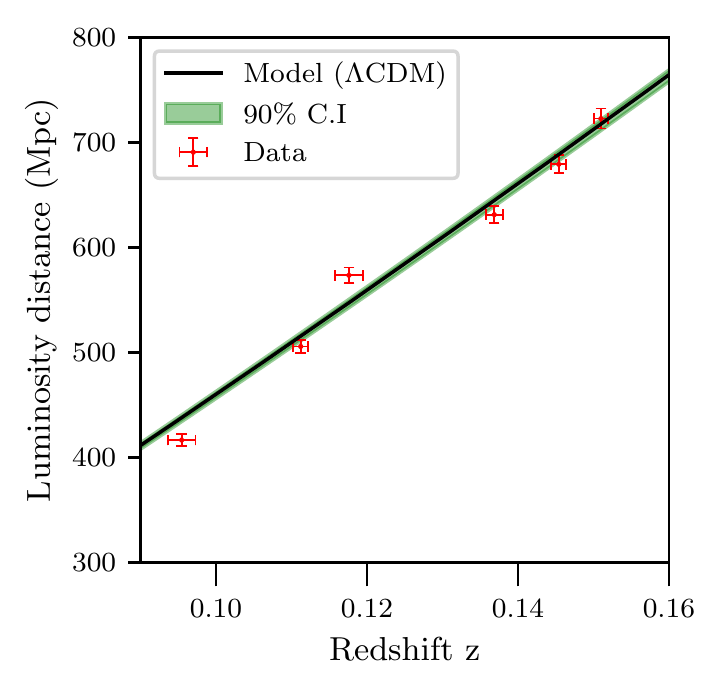}
\includegraphics[width = \linewidth]{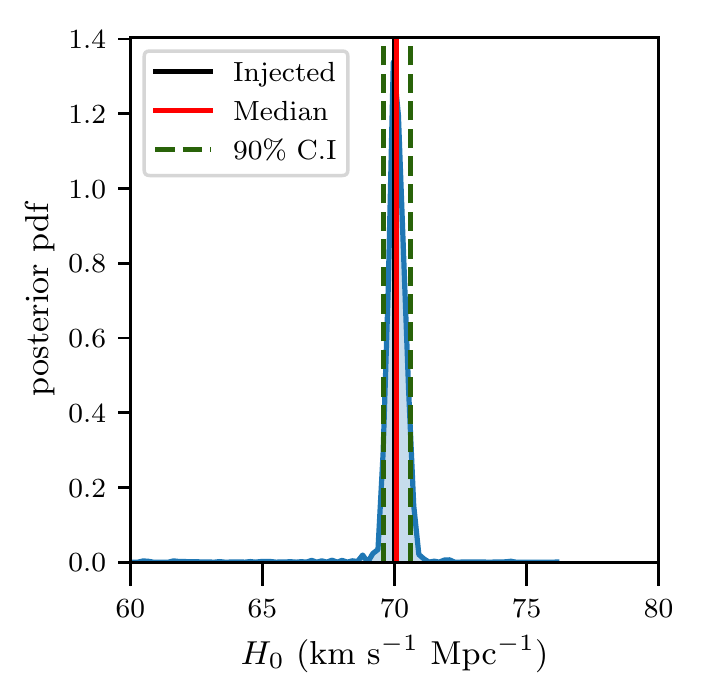}
\caption[Optional caption for list of figures]{{\it Top panel:} Luminosity distance (Mpc) plotted against the inferred redshifts, just as in Fig.~\ref{hubble1}, but now for the 5000 simulated GW events with Gaussian sky-localization errors discussed in Appendix~\ref{sec:app2}. The inferred redshift distributions are shown in Fig. \ref{corr}. {\it Bottom panel:} The posterior probability density for $H_0$ corresponding to the {\it top panel}. The prior on $H_0$ has been taken as uniform in the interval $[50,90]~ \kmspMpc$. The black line is the injected value $H_0 = 70 \kmspMpc$ used for the source simulations. The red line is the median, which coincides with the injected value. The green dashed lines mark the boundary of the $90\%$ confidence interval.}
\label{hubble}
\end{figure}

The corresponding Hubble plot is shown in the top panel of Fig.~\ref{hubble} while the posterior distribution of $H_0$ computed from it is shown in the bottom panel. The posterior is computed using the Bayesian technique as described in Sec.~\ref{sec4a}, where the prior on $H_0$ is taken to be uniform in the interval $[50,90] ~ \kmspMpc$.  The constraints on $H_0$ are obtained as $H_0 = 70.07^{+0.53}_{-0.49} \kmspMpc$, which defines the $90\%$ confidence interval around the median value $70.07 \kmspMpc$.

\section{The inferred vs true redshift distributions of the simulated GW sources}

\begin{figure*}
\centering
 \includegraphics[width = \textwidth, 
 height = 0.8\textwidth]{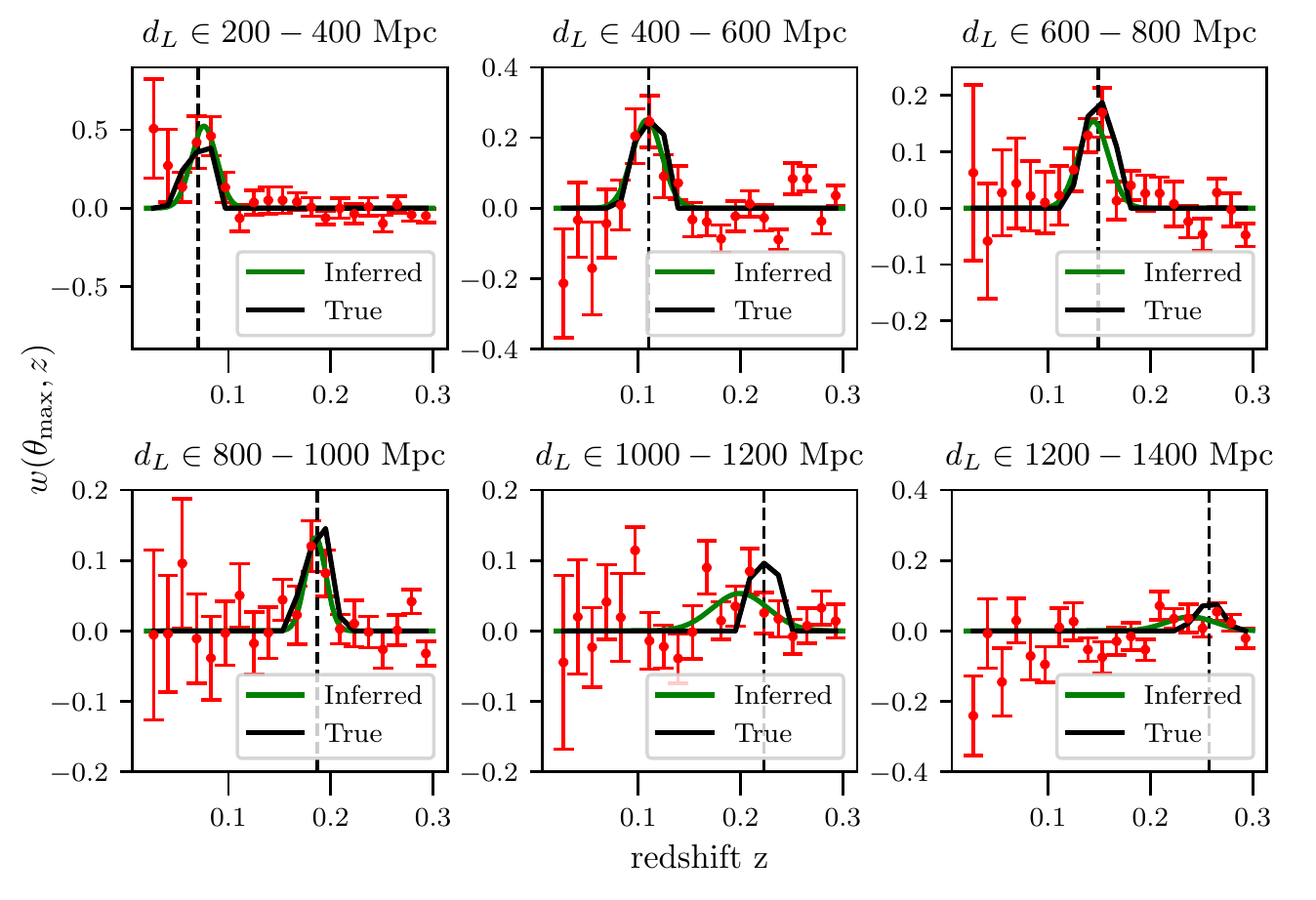}
\caption{Redshift distribution of the 5100 GW sources as obtained from the angular cross-correlation. We also show the true redshift distribution of the sources as obtained from the simulation. The black curve in each $d_{\rm L}$ bin represents the true redshift distribution whereas the green curve is the inferred distribution assuming a Gaussian form. The black dashed line in each bin shows the true average redshift of the sources in a given luminosity distance bin.}
\label{truersd}
\end{figure*}

In sec. \ref{sec2b}, we discussed the method of recovering the true redshift distribution of the GW sources from the computed cross-correlation. We argued that the assumption of a simple Gaussian cross-correlation (Eq. \eqref{gaussian}) is sufficient for our purpose, since in our simulation we have chosen the galaxy bias to be very close to unity. It is good to check how well this assumption holds true in our case. In this section, we show, along with the previously computed redshift distributions, the true redshift distributions of the GW sources. This has been plotted in Fig. \ref{truersd}. The true redshifts of the GW sources are computed directly from their true luminosity distances assuming a flat $\Lambda$CDM cosmology with $H_0 = 70 ~ \kmspMpc$. The corresponding redshift distribution is shown in thick black curve. The green curve is the inferred redshift distribution. The black dashed line is the true average redshift of the GW sources in each bin. As can be seen, the inferred redshift distribution does not differ much from the true distribution for closer sources (luminosity distances upto $\sim 1000$ Mpc). For sources with $d_{\rm L} > 1000$ Mpc, there is a slight discrepancy, which might be due to poor SNR at relatively large distances.

\bibliography{main.bib}
\bibliographystyle{apj.bst}
\end{document}